\documentclass[10pt,journal,a4paper,final,oneside,twocolumn]{IEEEtran}
\IEEEoverridecommandlockouts
\usepackage{cite}
\usepackage{float}
\usepackage{stfloats}
\usepackage{fancyhdr}
\usepackage{color}
\usepackage{amsmath}
\bstctlcite{IEEEexample:BSTcontrol}
\usepackage{graphicx} 
\graphicspath{{./figures/}}%说明待加载的图片的路径，在后期图片加载的时候不用再写路径，直接用图片名称
\usepackage{algorithm}
\usepackage{algorithmic}
\usepackage{soul, color, xcolor}

\usepackage{amssymb}
\usepackage{epsfig}
\usepackage{epstopdf}
\usepackage{subfigure}
\usepackage{txfonts}
\usepackage{bm}
\begin{document}
%	\pagenumbering{number}
	\title{Robust NOMA-assisted OTFS-ISAC Network Design with 3D Motion Prediction Topology
	}
	
        \author{Luping Xiang, {\em Member, IEEE}, Ke Xu,  {\em Student Member, IEEE}, Jie Hu, {\em Senior Member, IEEE}, Christos Masouros, {\em Senior Member, IEEE}, Kun Yang, {\em Fellow, IEEE}
        \thanks{Luping Xiang, Ke Xu and Jie Hu are with the School of Information
        and Communication Engineering, University of Electronic Science and
        Technology of China, Chengdu 611731, China, email: luping.xiang@uestc.edu.cn, 202121010634@std.uestc.edu.cn, hujie@uestc.edu.cn. }
        \thanks{Christos Masouros is with the Department of Electronic and Electrical
        Engineering, University College London, London WC1E 7JE, U.K., email: chris.masouros@ieee.org.}
        \thanks{Kun Yang is with the School of Computer Science and Electronic Engineering,
        University of Essex, Essex CO4 3SQ, U.K., e-mail: kunyang@essex.ac.uk.}
        %\thanks{Lajos Hanzo is with the School of Electronics and Computer Science, University of
        %Southampton, Southampton SO171BJ, U.K., e-mail: lh@ecs.soton.ac.uk}
        }
	%\textit{(Corresponding author: Jie Hu.)}
	\maketitle
	
	\thispagestyle{fancy} % IEEE模板在\maketitle后会自动声明\thispagestyle{plain}，
	% 导致第一页什么都没有。所以得把plain更改为fancy
	\lhead{} % 页眉左，需要东西的话就在{}内添加
	\chead{} % 页眉中
	\rhead{} % 页眉右
	\lfoot{} % 页眉左
	\cfoot{} % 页眉中
	\rfoot{\thepage} %页眉右，\thepage 表示当前页码
	\renewcommand{\headrulewidth}{0pt} %改为0pt即可去掉页眉下面的横线
	\renewcommand{\footrulewidth}{0pt} %改为0pt即可去掉页脚上面的横线
	\pagestyle{fancy}

    \rfoot{\thepage} % 页眉右

	\begin{abstract}
This paper proposes a novel non-orthogonal multiple access (NOMA)-assisted orthogonal time-frequency space (OTFS)-integrated sensing and communication (ISAC) network, which uses unmanned aerial vehicles (UAVs) as air base stations to support multiple users. By employing ISAC, the UAV extracts position and velocity information from the user's echo signals, and non-orthogonal power allocation is conducted to achieve a superior achievable rate. A 3D motion prediction topology is used to guide the NOMA transmission for multiple users, and a robust power allocation solution is proposed under perfect and imperfect channel estimation for Maxi-min Fairness (MMF) and Maximum sum-Rate (SR) problems. Simulation results demonstrate the superiority of the proposed NOMA-assisted OTFS-ISAC system over other systems in terms of achievable rate under both perfect and imperfect channel conditions with the aid of 3D motion prediction topology. 
	%	With the development of wireless communication, a large number of network nodes have brought huge energy supply problems. 

    \end{abstract}
	\section{Introduction}
%\IEEEPARstart{W}{ith} the next generation wireless network development, the  $7544$ billion in 2025\cite{IOT}.
%, causing many derived communication problems in numerous scenarios~\cite{6Gsurvey}.%

%Especially the high quality wireless communication for high-mobility vehicular networks has attracted more and more attention. The radar sensing aided communication system is considered a visionary solution.
%With the development of electronic components and the requirement of the communication, higher frequency is achieved to realise the gratifying service, which has already occupied some spectral resource of the radar.  The earliest concept of the ISAC is proposed in 1960 \cite{earliesISAC}. Since 1990, this technology had been greatly iterated, which is motivated by the technology of MIMO. The emergence of MIMO communication and MIMO radar greatly improve the communication rate, degree of freedom (DoFs) and sensing  
%accuracy of radar, which make the ISAC have practical value \cite{2004radar,MIMOradar,MIMIsystem}. 

Numerous digital devices in 6G will result in communications in higher frequencies, motivating the design of integrated sensing and communication (ISAC) technology~\cite{IOT,ISAC_application}. The potential of ISAC technology is evident in its applicability to vehicular communication, environmental surveillance, urban digital infrastructure, and human-machine interfaces \cite{Application1,Application2}.
By embedding  information into radar pluses, the basic function of ISAC was first accomplished in \cite{earliesISAC}. Advancements in hardware and signal processing techniques greatly improve the communication rate, degree of freedom (DoFs), and sensing  
accuracy of radar \cite{2004radar,MIMOradar,MIMIsystem}. 
%In \cite{MIMO_interfernce1,MIMO_interfernce2},  mutual interference (MI) in the co-design of MIMO communication and MIMO radar is addressed.
% Recent researches focus on designing novel dual-functional waveforms to satisfy the requirement of communication and radar. 

Due to the high spectral efficiency and multipath fading resistance, orthogonal frequency division multiplexing (OFDM) is intensively investigated in  ISAC  with an emphasis on radar imaging, target identification, etc.~\cite{OFDM_ISAC,OFDM_radar,OFDM_inter}.  However, in  high-mobility scenarios, OFDM waveform suffers from substantial Doppler offset \cite{OTFSperformance}. Fortunately, a new waveform design, known as orthogonal time frequency space (OTFS)~\cite{2017OTFS}, was established in the delay-Doppler (DD) domain to deal with 
Doppler offset \cite{OTFS_OFDM}.
The parameters within the DD domain inherently correlate with the spatial position and velocity of reflectors, rendering it well suited for radar-based sensing. As a result, OTFS emerges as a potential waveform of choice for the integrated sensing and communication (ISAC) framework. % [ref] there are many related works about the OTFS-ISAC. 
Single-antenna and multiple-antenna  OTFS-ISAC were proposed in \cite{OTFSISACsignal} and \cite{OTFSISACMIMO}, respectively, demonstrating   superior rate and estimation accuracy  over OFDM-ISAC.
Inspired by the conventional OFDM, the OTFS sensing in the time-frequency(TF) domain is proposed in \cite{OTFSradar_IOT}, where the DD profile is obtained through Fourier transform. A low-complexity matched-filter (MF) algorithm in the DD domain is proposed to estimate the distance and velocity in the ISAC system \cite{OTFSradar}. Considering more practical scenarios, an iterative optimization algorithm was proposed to deal with continuous delay and Doppler estimation for the OTFS-ISAC signal over the multipath channel \cite{robust-OTFS-ISAC}.
In \cite{multiUser},  orthogonal resource allocation is considered in ISAC for multiple users to maximize the estimation accuracy while guaranteeing the communication quality of service (QoS). An OTFS-ISAC transmission methodology incorporating a roadside unit (RUS) has been introduced for multi-vehicle scenarios \cite{YYJOTFSISACVhi}. Following the RUS's estimation of a vehicle's position and velocity, a vehicular topology is formulated in the adjacent lanes to facilitate the communication process.
% In the latest ISAC formwork enabled by intelligent reflecting surface (IRS) \cite{IRS-ISAC}\cite{IRS-ISAC2}, the IRS assists multiple users' communication by the  user's position in the three-dimensional space. 

% the user will receive the LOS signal marked by black solid  line and the NLOS signal marked by blue solid  line, where the BS receive the reflective LOS signal marked by black dotted line and the NLOS signal marked by  blue dotted line.For example, the signal is reflected from BS through Reflector:2 to user. Due to the curvature of the user's reflective surface, the signal reflected from user through Reflector:3 to BS.

\begin{figure}[t]
		\centering
		\includegraphics[width=\linewidth]{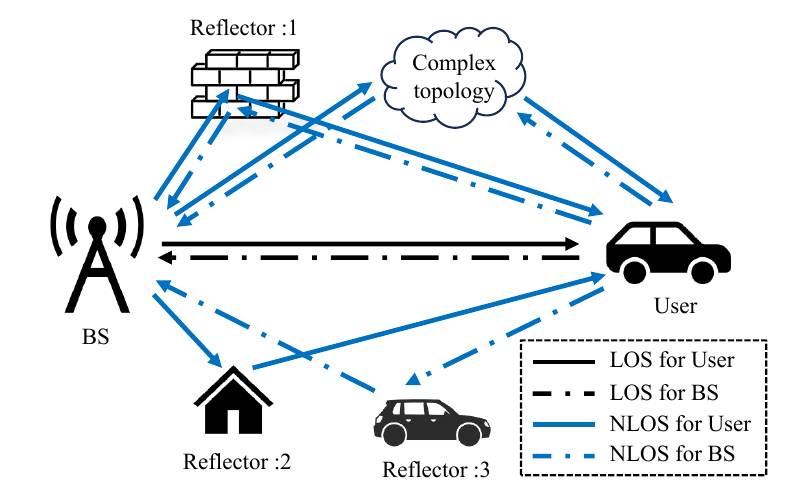}
		\caption{The LOS and NLOS path for the ISAC system.} \label{fig:LOS_NLOS}
	\end{figure}

The correlation between a user's position and velocity and the delay and Doppler of the OTFS channel offers an opportunity for the base station (BS) to facilitate users in avoiding the channel estimation process via pre-processing, as discussed in \cite{YYJOTFSISACVhi,Bypassing}. This strategy significantly streamlines the frame structure while minimizing pilot overhead. Nonetheless, it's imperative to note that this method is most effective within the confines of a line of sight (LOS) channel model. Despite the utility of radar sensing in obtaining LOS channel information, it falls short in accurately detecting non-line of sight (NLOS) paths. This lack of precision prevents the effective mitigation of NLOS impact via straightforward pre-processing. As depicted in Fig.~\ref{fig:LOS_NLOS}, the NLOS paths for users vary from those for the BS. This variance prevents the BS from accurately estimating the downlink NLOS channel by merely analyzing the return NLOS channel. This introduces the necessity to factor in the imperfect channel estimation of the NLOS path during pre-processing in order to address a more generalized channel model comprising both LOS and NLOS paths.

Additionally, leveraging prior knowledge, such as user location as perceived by the BS, can significantly refine non-orthogonal multiple access (NOMA) power distribution, thereby boosting communication throughput for multiple users. Such an approach obviates the necessity for users to transmit their positional information to the BS via an uplink procedure.  Recent advancements in NOMA-assisted ISAC research have opened new avenues in areas like beamforming design, interference elimination, and multi-user dynamics \cite{ISAC_NOMA,NOMA-ISAC-interference,OTFS_NOMA_Imperfect}. However, robust design remains an area for further exploration \cite{OTFS_NOMA_Imperfect}. Importantly, the imperfect channel estimation resultant from the NLOS may influence power allocation, a factor previously unaccounted for in NOMA-assisted ISAC studies.

Motivated by the pursuit of amplifying the sensing gain and elaborating on existing studies related to NLOS challenges, we introduce a novel NOMA-integrated OTFS-ISAC framework tailored for multi-user scenarios, exhibiting potential for deployment within robust, high-velocity mobile networks anchored on UAVs. Within this context, The UAV is regarded as an air BS, where the LOS path between the user and UAV can be guaranteed to attend in the system \cite{UAVLOS}. After the UAV obtains the user's position and velocity via the signal echo spread in the LOS channel, the 3D motion prediction topology is implemented to guide the NOMA transmission for multiple users.  In addition, the influence of imperfect channel estimation will be evaluated in two NOMA classic problems: max-min fairness (MMF) and maximum sum-rate (SR). The SR problem focuses on increasing the sum rate of the system, whereas the MMF problem ensures fairness between users.

Our novel  contributions are  explicitly contrasted  in Table I  and  are further summarised as follows: 
\begin{table*}[t]
	\centering
	\renewcommand{\arraystretch}{1.3} %设置行高
	\caption{CONTRASTING THE CONTRIBUTIONS OF THIS WORK TO THE LITERATURE}
	\label{table1} 
\begin{tabular}{c|c|c|c|c|c|c|c}
	\hline
	Contribution   & This work & \cite{robust-OTFS-ISAC}   & \cite{OTFSISACsignal},\cite{OTFSradar},\cite{YYJOTFSISACVhi}      & \cite{IRS-ISAC},\cite{IRS-ISAC2} &\cite{ISAC_NOMA},\cite{NOMA-ISAC-interference}   &\cite{NOMA_imperfect},\cite{OTFS_NOMAMA}&\cite{ISAC_NLOS }    \\ \hline
     Radar sensing &$\checkmark$ & $\checkmark$  &  $\checkmark$   &  $\checkmark$   & $\checkmark$      &    & $\checkmark$    \\ \hline
	OTFS  &$\checkmark$ &  $\checkmark$  &  $\checkmark$  &  &         & $\checkmark$  &\\ \hline
    NOMA for multiple user  & $\checkmark$&                &  & & $\checkmark$ & $\checkmark$       &       \\ \hline
	3D motion model    & $\checkmark$  &      &  & $\checkmark$ &           &    &        \\ \hline
	Multiple path for ISAC system  & $\checkmark$  &  $\checkmark$ &  & &  &          &     $\checkmark$                    \\ \hline
 Robust power allocation & $\checkmark$  &   &  & &  &          &                        \\ \hline
\end{tabular}\label{parametertable}
\end{table*}

\begin{itemize}
	\item We propose a  NOMA-assisted OTFS-ISAC system, where the UAV serves as the air BS to support multiple users. By employing ISAC, the UAV extracts the position and velocity information from the user's echo signals during communication.  On the UAV side, non-orthogonal power allocation is conducted based on the extracted information to achieve a superior data rate. 
	
	%scenarios. Using power allocation, the diverse requirements of the multiuser are guaranteed.In order to satisfy the realistic transimission scenarios, the non line of sight (NLOS) and line of sight (LOS) channel, that can't be sensed by the radar, is considered.
% 	\item For the sake of clarity, the protocol of our system is proposed to show our advantages over traditional system. Based on the speed, range and angle analyzed by radar, we proposed the method to establish the user's 3D motion topological. 
    \item Additionally, we examine a three-dimensional motion model, where the distance, velocity, and angle of the user are retrieved from echo signals.  \textcolor{black}{
    The above parameters can only describe the LOS channel between the UAV and the user, hence, the robust power allocation will be investigated with considering the impact of the NLOS channel.}
    
    % \textcolor{red}{The unknown NLOS channel will be considered in the power allocation to get higher date rate.}
	
	\item  We derive a closed-form solution to the MMF and SR  problem involving non-orthogonal power allocation in OTFS-ISAC systems.  Simulation results demonstrate the superiority of our proposed NOMA-assisted OTFS-ISAC system over the OMA-assisted OTFS-ISAC system in terms of MMF and SR.  
\end{itemize}
\begin{figure}[t]
		\centering
		\includegraphics[width=\linewidth]{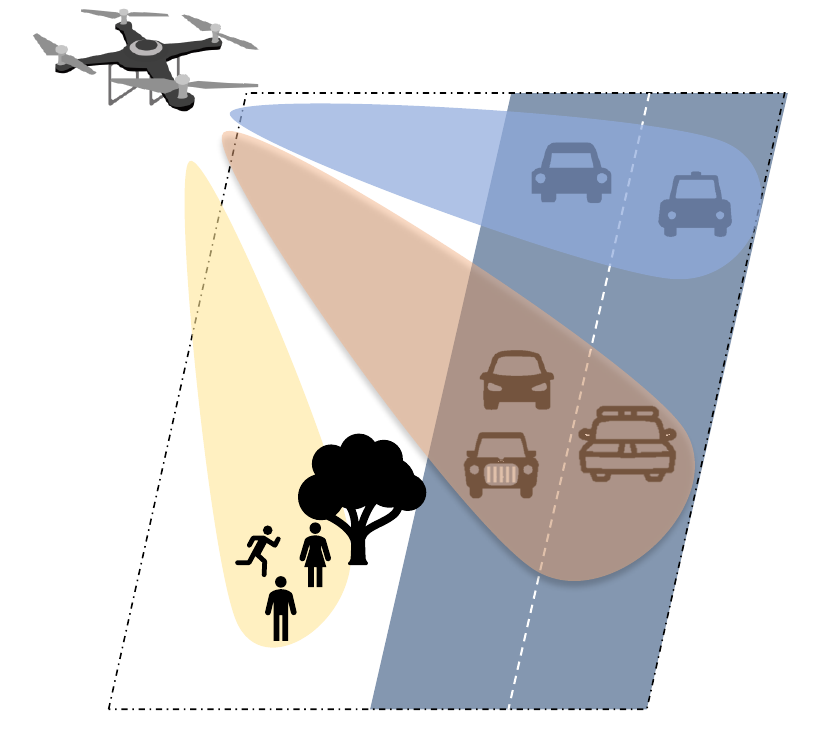}
		\caption{NOMA-assisted OTFS-ISAC networks} \label{fig:system}
	\end{figure}

 \begin{figure}[t]
		\centering
		\includegraphics[width=80mm]{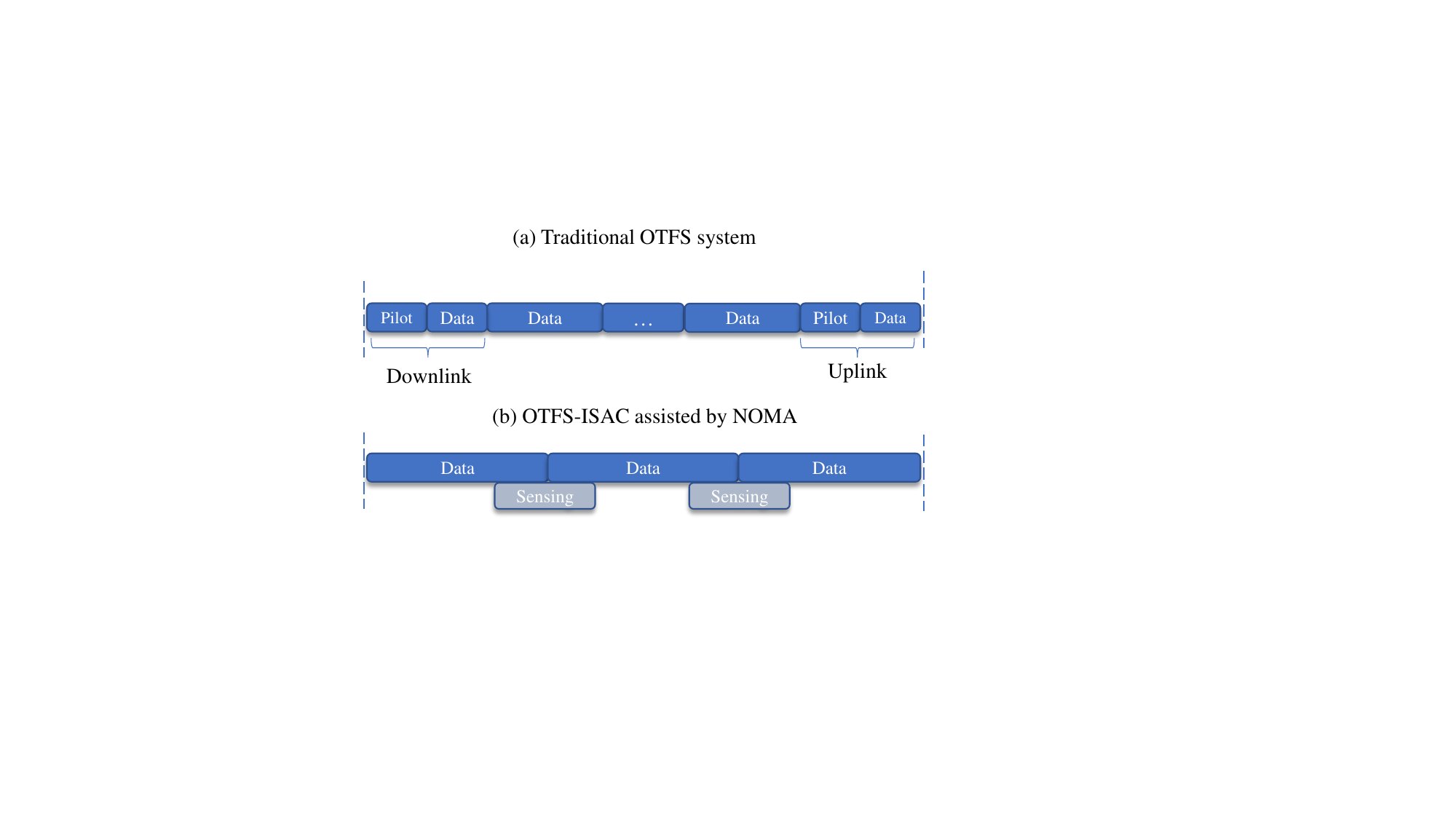}
		\caption{The comparison between the transmission protocol} \label{superiority}
\end{figure}

\section{OTFS-ISAC system assisted by NOMA}
The NOMA-assisted OTFS-ISAC network is shown in Fig.~\ref{fig:system}, where a UAV  supports $G$ clusters and the $g$-th cluster has $P_g$ users, where the $g \in \left\{1,2,\dots,G\right\}$. 
%The $U_{p,g}$  denotes the $p$-th user in the $g$-th cluster, where the $p \in \left\{1,2,\dots,P_g\right\}$.% 
We assume the UAV is equipped with 2 UPAs. One UPA at the UAV transmits the OTFS-ISAC signal to all clusters while the other receives the echo signal from the users. We presume that echo signals do not interact with one another. As shown in the Fig~\ref{fig:system}, the UAV performs beamforming for the following time slot after analyzing the users' motion parameters acquired from echoes in the preceding time slot. We assume that users  are autonomous and do not  block each other during the movement. 
%\subsection{Transmission Protocol}

The transmission protocol of the traditional OTFS communication system and the NOMA-assisted OTFS-ISAC system are contrasted in Fig.~\ref{superiority}  . Specifically, as illustrated in Fig.~\ref{superiority}, pilots are required to be transmitted before the data transmission in the conventional OTFS system. Additionally, the CSI obtained in the previous data frame would be outdated for the subsequent frame,  resulting in communication performance degradation.  %And the uplink stage with large scale  information transmitted by the user is exploited by the UAV to assist the transmission of NOMA. 
By  contrast, in  a NOMA-assisted OTFS-ISAC system, the UAV can obtain the position and velocity of the users through echoes at no additional cost. The UAV can obtain the OTFS channel by converting the position and velocity information into delay and Doppler information, the user can bypass the step of channel estimation by the UAV's pre-processing, resulting in pilot-free transmission. 
In addition, the large-scale fading inferred from the position can guide the power allocation of NOMA, which in turn can improve the data rate of the OTFS-ISAC system.

\subsection{OTFS-ISAC signal}
At the transmitter, we assume that the UAV transmits the OTFS-modulated symbol $x_{p,g}[k,l]$  in the DD domain to the $p$-th user in the $g$-th cluster  $U_{p,g}$, where $k=0, 1, \cdots, N-1$ and $l=0, 1, \cdots, M-1$ are the Doppler and delay indices, respectively. Here, $M$ and $N$ represent the total number of subcarriers and time slots, respectively. The  DD-domain signal is then converted  to the TF-domain using the inverse symplectic finite Fourier transform (ISFFT), which can be expressed as:
\begin{align}
	X_{p,g}{[n, m]}=&\frac{1}{\sqrt{N M}} \sum_{k=0}^{N-1} \sum_{l=0}^{M-1} x_{p,g}[k, l] e^{j 2 \pi\left(\frac{n k}{N}-\frac{m l}{M}\right)} ,
\end{align}
where $n=0, 1, \cdots, N-1$ and $m=0, 1, \cdots, M-1$ are the time and frequency indices in the TF-domain.

Invoking the ideal rectangular transmit pulse $g_{\rm{tx}}(t)$, the time-domain signal $X_{p,g}{[n,m]}$ is converted to the continuous waveform $s_{p,g}(t)$ by the Heisenberg transform, which is expressed as:
\begin{align}
	s^{p
	}_{g}(t)=&\sum_{n=0}^{N-1} \sum_{m=0}^{M-1} X_{p,g}[n, m] g_{\rm{tx}}(t-n T) e^{j 2 \pi m \Delta f(t-n T)}.\label{Heisenberg}
\end{align}

To serve $P_g$ users in the $g$-th cluster, the UAV transmits the superimposed signal $s_{g}(t)=\sum_{p=1}^{P_g}\omega_{p,g} s_{p,g}(t)$, where $\omega_{p,g}$ denotes the power assigned to the $p$-th user. The transmitted signal to $G$ clusters can be expressed as  $\mathbf{s}(t)={\left[s_{1}(t),s_{2}(t), \ldots,s_{G}(t)\right]}^{\text{T}}$. Considering the UPA with a size of $N_{x} \times N_{y}$, the steering vector $\mathbf{a}\left(\theta_g, \varphi_g\right) \in \mathbb{C}^{N_{x}  N_{y}\times 1}  $ can be defined as:
\begin{align}
    \begin{split}
   \mathbf{a}\left(\theta_{g}, \varphi_{g}\right)=&\frac{1}{\sqrt{N_{x} N_{y}}} \left[1, \ldots, e^{j \pi\sin \theta_{g}\left(n_{x} \sin \varphi_{g}+n_{y} \cos \varphi_{g}\right)}\right.,\\
   &\left.\ldots, e^{j  \pi \sin \theta_{g}\left(N_{x} \sin \varphi_{g}+N_{y} \cos \varphi_{g}\right)}\right]^{\text{T}},
   \end{split}
\end{align}
where the $\theta_g$ and $\varphi_g$ are the azimuth and elevation of the $g$-the cluster, respectively. Additionally, $n_x=1,2,\dots, N_x$ and $n_y=1,2,\dots,N_y$ are \textcolor{black}{the indices of the transmit antenna.}
Defining  $\mathbf{A}={\left[\mathbf{a}\left(\theta_1, \varphi_1\right),\dots,\mathbf{a}\left(\theta_G, \varphi_G\right)\right]}$, the transmitted signal can be formulated as:
\begin{align}
	{\bar{\mathbf{s}}}(t)=&\mathbf{A}\mathbf{s}(t) .
\end{align}

\subsection{\textcolor{black}{Radar Sensing Process}}

The UAV  receives the  echo  signal via the radar channel $\mathbf{H}_{p,g}(t, \tau)$, which can be expressed as:
\begin{align}
\begin{split}
	&\mathbf{H}_{p,g}(t, \tau)= \beta_{p,g}\mathbf{b}\left(\theta_{p,g},\varphi_{p,g}\right) \mathbf{b}^{\mathrm{H}}\left(\theta_{p,g},\varphi_{p,g}\right)\delta\left(t-\tau_{p,g}\right)\times\\& e^{j 2 \pi \nu_{p,g} t}+\sum_{i=1}^{N_{p,g}}
    \hat \beta_{p,g}^{\text{R},i}\mathbf{b}\left(\hat \theta_{p,g}^{\text{R},i},\hat \varphi_{p,g}^{\text{R},i}\right)\mathbf{b}^{\mathrm{H}}\left(\hat \theta_{p,g}^{\text{R},i},\hat \varphi_{p,g}^{\text{R},i}\right)\delta\left(t-\hat \tau_{p,g}^{\text{R},i}\right) e^{j 2 \pi \hat \nu_{p,g}^{\text{R},i} t}.\label{RadarChannel}
    \end{split}
\end{align}
where the  $\beta_{p,g}$, $\tau_{p,g}$ and $\nu_{p,g}$ respectively represent the reflection coefficient, delay and the Doppler offset of the LOS channel with the direction $\left(\theta_{p,g},\varphi_{p,g}\right)$ between the $p$-th user in the $g$-th cluster and the UAV. The  $\beta_{p,g}^{\text{R},i}$, $\tau_{p,g}^{\text{R},i}$ and $\nu_{p,g}^{\text{R},i}$ represent the reflection coefficient, delay and the Doppler offset of the $i$-th radar NLOS path with the direction $\left(\theta_{p,g}^{\text{R},i},\varphi_{p,g}^{\text{R},i}\right)$.

In Eq.\eqref{RadarChannel} $\mathbf{b}\left(\theta_{p,g},\varphi_{p,g}\right)$ is the receive steering vector, which can be expressed as:
\begin{align}
    \begin{split}
   \mathbf{b}\left(\theta_{p,g}, \varphi_{p,g}\right)=& \frac{1}{\sqrt{N_{x}N_{y}}} \left[1, \ldots, e^{j  \pi\sin \theta_{p,g}\left(n_{x} \sin \varphi_{p,g}+n_{y} \cos \varphi_{p,g}\right)}\right.,\\
   &\left.\ldots, e^{j  \pi\sin \theta_{p,g}\left(N_{x} \sin \varphi_{p,g}+N_{y} \cos \varphi_{p,g}\right)}\right]^{T}.
   \end{split}
\end{align}
The $\mathbf{b}\left(\hat \theta_{p,g}^{\text{R},i},\hat \varphi_{p,g}^{\text{R},i}\right)$ will be obtained by replacing the  $\theta_{p,g}$ and $\varphi_{p,g}$ in $\mathbf{b}\left(\theta_{p,g}, \varphi_{p,g}\right)$ with $\hat \theta_{p,g}^{\text{R},i}$ and $\hat \varphi_{p,g}^{\text{R},i}$.

Furthermore, the echo  signal of $U_{p,g}$ can be formulated as:
\begin{align}
  \begin{split}
    &\mathbf{r}_{p,g}(t)=\beta_{p,g}\mathbf{b}\left(\theta_{p,g}, \varphi_{p,g}\right) \mathbf{b}^{\mathrm{H}}\left(\theta_{p,g}, \varphi_{p,g}\right) \mathbf{a}\left(\theta_g, \varphi_g\right)\times\\&  s_{g}(t-\tau_{p,g}) e^{j 2 \pi \nu_{p,g} t}+\sum_{i=1}^{N_{p,g}}
    \hat \beta_{p,g}^{\text{R},i}\mathbf{b}\left(\hat \theta_{p,g}^{\text{R},i},\hat \varphi_{p,g}^{\text{R},i}\right)\mathbf{b}^{\mathrm{H}}\left(\hat \theta_{p,g}^{\text{R},i},\hat \varphi_{p,g}^{\text{R},i}\right)\\&\times\mathbf{a}\left(\theta_g, \varphi_g\right)s_{g}\left(t-\hat \tau_{p,g}^{\text{R},i}\right) e^{j 2 \pi \hat \nu_{p,g}^{\text{R},i} t}+
    \mathbf{z}(t),
    \end{split}
\end{align}
where the $\mathbf{z}(t) $ is the white Gaussian noise.

To facilitate communication, the channel parameters can be obtained by following steps.
First, the angle $\left(\theta_{p,g}, \varphi_{p,g}\right)$ and $\left(\hat \theta_{p,g}^{\text{R},i},\hat \varphi_{p,g}^{\text{R},i}\right)$  can be estimated by using a mature method called MUSIC \cite{MUSIC}, which has great efficiency and high resolution. Then, the  echo  signal  without angle information can be expressed as:
\begin{align}
  \begin{split}
\bar{r}_{p,g}(t) =\beta_{p,g} s_{g}(t-\tau_{p,g}) e^{j 2 \pi \nu_{p,g} t}+\sum_{i=1}^{N_{p,g}}
    \hat \beta_{p,g}^{\text{R},i}s_{g}(t-\hat \tau_{p,g}^{\text{R},i}) e^{j 2 \pi \hat\nu_{p,g}^{\text{R},i} t}.
   \end{split}
\end{align}
Second, the UAV performs MF  on the echo signal to obtain $\tau_{p,g}$, $\nu_{p,g}$, $\hat \tau_{p,g}^{\text{R},i}$ and $ \hat\nu_{p,g}^{\text{R},i}$. The 
correlated value function $\jmath(\tau, \nu)$ can be represented as follows:
\begin{align}
    \jmath (\tau, \nu)=\int_{0}^{\Delta T}\bar{r}_{p,g}(t) {{s}_{g}}^{*}(t-\tau) e^{-j 2 \pi \nu t} \mathrm{~d} t ,
\end{align}
where $\Delta T$ represents the frame time duration, and $*$ represents the conjugate operator.  Although, both the radar's LOS and NLOS channel information can be obtained  by the radar sensing process, only the LOS channel of radar is  highly correlated to the LOS channel of communication, which can be applied in the
communication pre-processing. \textcolor{black}{ The NLOS channel sensed by the radar is different from the NLOS channel in the communication. But, the NLOS path sensed by the radar 
 can describe the complexity of the environment \cite{K_factor}, where we define $e_{p,g}$ to represent the strength of the NLOS channel in the environment:}
\begin{align}
e_{p,g}=\frac{\sum_{i=1}^{N_{p,g}}
   \left( \hat \beta_{p,g}^{\text{R},i}\right)^2 }{\left(\beta_{p,g}\right)^2}.   
    \end{align}
The estimated of $e_{p,g}$ can be obtained by the function $ \jmath (\tau, \nu)$:
\begin{align}
\hat e_{p,g}=\frac{\sum_{i=1}^{N_{p,g}}
   \left(\jmath (\hat\tau_{p,q}^{\text{R},i}, \hat\nu_{p,q}^{\text{R},i})\right)^2 }{\left(\jmath (\tau_{p,q}, \nu_{p,q})\right)^2},   
\end{align}
which will be considered in the following NOMA power allocation.

\subsection{Communication Process}

The communication channel is different from the radar channel, which is consisted by multiple paths from UAV to the user, with the LOS path predominating. The communication channel between the $p$-th user in the $g$-th cluster and the UAV can be expressed as:
\begin{align}
%	\begin{split}
        &{\mathbf{\bar H}}_{p,g}\left(t,\tau\right)=h_{p,g} \mathbf{b}^{\text{H}}\left(\theta_{p,g}, \varphi_{p,g}\right)\delta\left(t-\frac{\tau_{p,g}}{2}\right) e^{j 2 \pi \nu_{p,g} t}
        \nonumber\\&+ \sum_{i=1}^{N_{p,g}}\hat h_{p,q}^{\text{C},i} \mathbf{b}^{\text{H}}\left(\hat \theta_{p,g}^{\text{C},i}, \hat \varphi_{p,g}^{\text{C},i}\right)\delta\left(t-\hat{\tau}_{p,q}^{\text{C},i}\right) e^{j 2 \pi \hat \nu_{p,q}^{\text{C},i} t}.
%	\end{split}
\end{align}
where  $h_{p,g}$ and $\hat h_{p,q}^{\text{C},i}$ represent the large scale loss of the LOS and NLOS, respectively. Additionally, $\left(\hat \theta_{p,g}^{\text{C},i}, \hat \varphi_{p,g}^{\text{C},i}\right)$ represents the receive direction of $i$-th NLOS, whereas $\hat{\tau}_{p,q}^{\text{C},i}$ and $\hat \nu_{p,q}^{\text{C},i}$ represent the $i$-th NLOS's delay and Doppler offset, respectively. 
Consequently, the received signal is expressed as:
\begin{align}
	\begin{split}
        y_{p,g}(t)={\mathbf{\bar H}}_{p,g}\left(t,\tau\right) \mathbf{ a}\left(\theta_{g}, \varphi_{g}\right)s_{p,g}(t).
	\end{split}
\end{align}
As soon as $y_{g,p}(t)$ is received,  Wigner transform is performed to  translate the time-domain signal to the TF domain.
\begin{align}
%\begin{split}
Y_{p,g}[n,m]
=&\int_{t^{\prime}} g_{\text{rx}}\left(t^{\prime}-t\right) y_{p,g}\left(t^{\prime}\right) \nonumber \\
&\times e^{-j 2 \pi f\left(t^{\prime}-t\right)} d t^{\prime}|_{t=n T, f=m \Delta f},
%\end{split}
\end{align}
where $g_{\text{rx}}\left(t\right) $ is the ideal rectangular pulse. Then, the Symplectic finite Fourier transform (SFFT) is applied to the discrete signal $Y_{p}[n,m]$ to obtain the information $y_{p}[k, l]$ in the DD domain.
\begin{align}
	%\begin{split}
		 y_{p,g}[k, l]=\frac{1}{\sqrt{N M}} \sum_{n=0}^{N-1} \sum_{m=0}^{M-1} {Y_{p,g}}[n, m] e^{-j 2 \pi\left(\frac{n k}{N}-\frac{m l}{M}\right)}.
	%\end{split}
\end{align}

\subsection{Three-dimensional motion topology}
\begin{figure}[H]
		\centering		\includegraphics[width=0.90\linewidth]{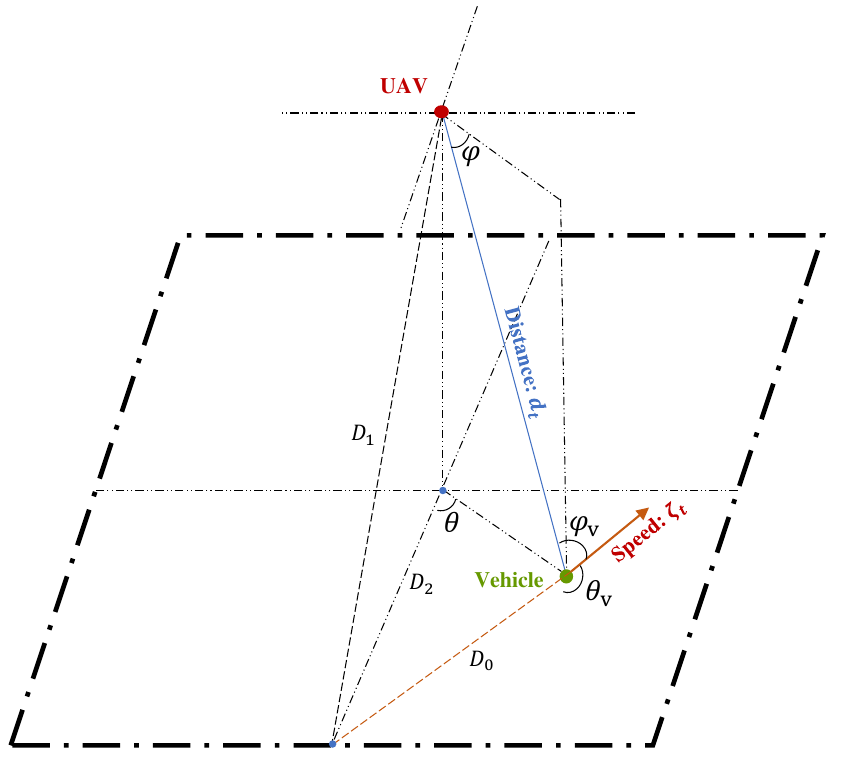}
		\caption{The 3-D user movement model} \label{motion}
\end{figure}
In our proposed NOMA-assisted OTFS-ISAC system, the UAV  assists the  OTFS-ISAC signal transmission by estimating the user's position in the next time slot. Hence, a 3D motion model is introduced to improve  estimation precision of the  user's position in the subsequent time slot.
Fig.~\ref{motion} depicts a schematic of topological 3D motion, where the UAV estimates relavant  parameters of the users on the ground. We assume that the angle of the velocity $\theta_v$ has been derived from the user position in the previous time slot. At the time slot $t$, the distance between the user and the UAV is $d_t$, while the azimuth and elevation angle are $\theta$ and $\varphi$, respectively. The distance $d_{t}$ can be calculated using the delay $\tau_{t}$ and the velocity of light $c$:
\begin{align}
	\begin{split}
		 d_{t}=\frac{c \tau_{t}}{2}.
	\end{split}
\end{align}

According to the geometric relationship, the angle $\varphi_{v}$ between the line connecting UAV and the user and the speed is given by 
\begin{align}
	\begin{split}
	\varphi_v=\pi-
	\arccos{\left(\frac{{d_{t}}^2+{\left(D_0\right)}^2-\left(D_1\right)^2}{2 d_{t} D_0}\right)},
	\end{split}
\end{align}
where we have:
\begin{align}
	\begin{split}
        D_0=&\frac{d_{t} \sin \theta \cos\varphi  }{\sin\left(\pi-\theta_{v}\right)}, \\
        D_1=&\sqrt{\left(d_{t}\sin \varphi\right)^2+\left(D_2\right)^2 },\\ D_2=&\frac{\sin\left(\pi-\theta_{v}\right)d_{t} \cos\varphi}{\sin\left(\theta_{v}-\theta\right)},
	\end{split}
\end{align}
where $D_0$, $D_1$ and $D_2$ represent the actual line segments that have been labelled in Fig.~\ref{motion}.
Furthermore, the user velocity   $\zeta_{t}$ can be derived from  $\varphi_v$ and  $\nu_{t}$ in the time slot $t$:
\begin{align}
	\begin{split}
	\zeta_{t}=\frac{c \nu_{t} }{\cos \varphi_{v} f_c},
	\end{split}
\end{align}
where   $f_c$ is the carrier frequency.
According to the 3D model, we may deduce the distance $d_{t+1}$ between the UAV and  the user at the next time slot $t+1$:
\begin{align}
	\begin{split}
		 d_{t+1}=\sqrt{{\left(d_{t}+\zeta_{t}T_{\text{otfs}}-d_{t}\sin\phi\right)}^2+{\left(d_{t}\sin\phi\right)}^2}.
	\end{split}
\end{align}
\begin{align}
	\begin{split}
	\phi=
	\arccos{\left(\frac{{D_{0}}^2+{\left(D_1\right)}^2-\left(d_{t_0}\right)^2}{2 D_{0} D_1}\right)}.
	\end{split}
\end{align}
The $T_{\text{otfs}}$ represents the estimation interval for each place.

\section{Power allocation for the perfect and imperfect channel}
In this section, we present a power allocation algorithm for NOMA under perfect and imperfect channel assumptions in the context of two classic NOMA problems, namely MMF and SR. The LOS path information is inferred from   users' position and speed, which can be obtained via our proposed 3D motion model. The perfect channel assumption considers only the LOS path, whereas the imperfect channel assumption accounts for both LOS and NLOS paths.
%By employing NOMA, we enhance the communication rate and reduce inter-user interference. 
Classic NOMA strategies often employ user-pairing to maintain tractable complexity, as highlighted in \cite{MMF,MMNOMA}. Moreover, methodologies for multi-user pairing have received substantial attention, as delineated in \cite{mutilUser1,mutilUser2}. In the context of this paper, and without compromising generality, we delve into power distribution post-user-pairing, specifically for a dual-user setup. Our proposed framework exhibits scalability for accommodating a broader user base by employing existing pairing techniques. The power allocated to User 1 (U1) and User 2 (U2) is denoted as $\omega_1$ and $\omega_2$, respectively. The achievable rates achieved by U1 and U2 with successive interference cancellation (SIC) are denoted as $R_1$ and $R_2$, respectively.
% \subsection{Typical problems of NOMA}
% NOMA aims to realize the flexible power management to achieve the different purposes of the communication. Maximin fairness (MMF), sum rate (SR) and energy efficiency (EE) are three typical optimization problems in the NOMA power allocation. 

% First, the MMF problem provide fairness for all users. The all users' lower bound of communication performance is maximized to ensure no users will be abandoned. The MMF problem can be formulated as:
% \begin{align}
% 	\begin{split}
% \max~\min~\left\{R_{1}, R_{2}\right\}
% 	\end{split}
% \end{align}

% Second, the SR is the overall performance index of communication. The SR maximization is given by:
% \begin{align}
% 	\begin{split}
%         \max~\left(R_{1}+R_{2}\right)
% 	\end{split}
% \end{align}

% Third, the fascinating purpose in NOMA power allocation is to increase the EE. The EE maximization can be express as:
% \begin{align}
% 	\begin{split}
%         \max~\frac{\left(R_{1}+R_{2}\right)}{P_{\text{c}}+\left(\omega_{1}+\omega_{2}\right)}
% 	\end{split}
% \end{align}
% $P_{\text{c}}$ represent the power consumption in the actual transmission process.

\subsection{Perfect Channel}
Assuming perfect channel conditions, the communication channel is modeled as a point-to-point system dominated by LOS transmission, ignoring NLOS effects \cite{LOS}. The large-scale fading factor, $h_p$, represents the path loss between the UAV and the $p$-th user for any given cluster, where $\left|h_{1}\right|^{2} \le \left|h_{2}\right|^{2}$. Additionally, the distance between the UAV and $p$-th user, denoted by $d_{p}$, is defined. The angle-dependent differences in $h_p$ for U1 and U2, which are illuminated by a single beam, can be neglected. As a result, the expression for $h_p$ for U1 and U2 can be defined as:
\begin{align}
\begin{split}
h_p=\frac{G_{\text{T}} G_{\text{R}} \lambda^{2}}{(4 \pi)^{2} {d_p}^{2}},
\end{split}
\end{align}
where $G_{\text{T}}$ and $G_{\text{R}}$ represent the transmit gain and receive gain, respectively, and $\lambda$ denotes the wavelength of   electromagnetic waves.
The achievable rate of U1 and U2 using NOMA is formulated as follows:
\begin{align}
\begin{split}
R_{1}=\log {2}\left(1+\frac{\omega_{1}\left|h_{1}\right|^{2}}{\omega_{2}\left|h_{1}\right|^{2}+n_0}\right),
\end{split}
\end{align}
and
\begin{align}
\begin{split}
R_{2}=\log {2}\left(1+\frac{\omega_{2}\left|h_{2}\right|^{2}}{n_0}\right).
\end{split}
\end{align}
% Let $R_{2\rightarrow 1}$ represents U2 decode  U1's information, $R_{2\rightarrow 1}=\log _{2}\left(1+\frac{\omega_{1}\left|h_{2}\right|^{2}}{\omega_{2}\left|h_{2}\right|^{2}+n_0}\right)$. Due to the previous assumption $\left|h_{1}\right|^{2} \le \left|h_{2}\right|^{2}$, $R_{2\rightarrow 1} \ge R_{1}$ can be guaranteed, which indicating  that U2 with the strong channel can demodulate the information of U1 to removed the interference.
\subsubsection{Maximin Fairness}
In order to ensure fairness among different users, the MMF problem is introduced. Mathematically, this problem can be formulated as 
\begin{align}
	(P1):\mathop{\max\min}\limits_{\omega_1,\omega_2}\quad&\left\{R_{1}, R_{2}\right\}\label{per_MMF} \\
	s.t.\quad&\omega_{1} + \omega_{2} \le P_{\text{t}}, \tag{\ref{per_MMF}{a}}
\end{align}
The aim of this problem is to maximize the rate of the minimum rate user, thus promoting fairness among users. The optimal power allocation for U2 in the MMF problem can be obtained as $\omega_{2,\text{MMF}}^{*}=\frac{-\left(\left|h_{1}\right|^{2}n_0+\left|h_{2}\right|^{2}n_0\right)+\sqrt{\left(\left|h_{1}\right|^{2}n_0+\left|h_{2}\right|^{2}n_0\right)^{2}+4P\left|h_{1}\right|^{4}\left|h_{2}\right|^{2}n_0}}{2\left|h_{1}\right|^{2}\left|h_{2}\right|^{2}}$. The value of $\omega_{1,\text{MMF}}^{}$ is then determined as $\omega_{1,\text{MMF}}^{*}=P_{\text{t}}-\omega_{2,\text{MMF}}^{*}$. The proof could be found in \cite{MMF}.

\subsubsection{Sum-Rate}
The primary goal of SR is to optimize the rate while adhering to the constraints of the quality of service. This optimization problem is expressed as (P2), where the objective is to maximize the sum of R1 and R2:
\begin{align}
	(P2): \mathop {\max }\limits_{\omega_1,\omega_2} \quad 
	&R_1+R_2,\label{X} \\
	\textrm{s.t} \quad 
	& R_1\ge R_{1,\text{min}}, \tag{\ref{X}{a}}\\
	& R_2\ge R_{2,\text{min}}, \tag{\ref{X}{b}}\\
	&\omega_{1} + \omega_{2} \le P_{\text{t}}. \tag{\ref{X}{c}}
\end{align}
The optimization problem above is subject to the constraints \ref{X}(a)-\ref{X}(c), which require $R_1$ and $R_2$ to be higher than or equal to their respective minimum required rates, and the total power transmitted by U1 and U2 to be less than or equal to $P_t$. In this problem, $R_{1,\text{min}}$  and $R_{2,\text{min}}$  represent the minimum required rates for U1 and U2, respectively. By fully utilizing the transmit power, we  set $\omega_{1}  = P_{\text{t}}- \omega_{2}$. The optimization function can be expressed $f_{\text{SR}}(\omega_2)=R_1+R_2$, where the derivative function $f’_{\text{SR}}(\omega_2)$ is expressed as
\begin{align}
    {f’_{\text{SR}}}(\omega_2)=\frac{\left(\left|h_{2}\right|^{2}-\left|h_{1}\right|^{2}\right)n_0^{2}}{\left(\omega_2\left|h_{2}\right|^{2}+n_0\right)\left(\omega_2\left|h_{1}\right|^{2}+n_0\right)n_0}.
\end{align}
Under the condition   $\left|h_{1}\right|^{2} \le \left|h_{2}\right|^{2}$,   $f'_1(\omega_2)$ is always positive, which indicates that the optimal solution is obtained at the upper bound of $\omega_2$. In order to meet the constraints \ref{X}(a) and \ref{X}(b), the upper and lower bounds of $\omega_2$ are calculated as $\frac{P_{\text{t}}\left|h_{1}\right|^{2}-\left(2^{R_{1,\text{min}}}-1\right)n_0}{\left(2^{R_{1,\text{min}}}-1\right)\left|h_{1}\right|^{2}+\left|h_{1}\right|^{2}}$ and $\frac{\left(2^{R_{2,\text{min}}}-1\right)n_0}{\left|h_{2}\right|^{2}}$, respectively. Therefore, the optimal power allocation for U2 in the SR problem is given by $\omega_{2,\text{SR}}^{*}=\frac{P_{\text{t}}\left|h_{1}\right|^{2}-\left(2^{R_{1,\text{min}}}-1\right)n_0}{\left(2^{R_{1,\text{min}}}-1\right)\left|h_{1}\right|^{2}+\left|h_{1}\right|^{2}}$, and the corresponding optimal power to be allocated to U1 is $\omega_{1,\text{SR}}^{*}=P_{\text{t}}=\omega_{2,\text{SR}}^{*}$.

\subsection{Imperfect Channel}
In practical scenarios, the identification of the LOS channel from the echo signal is possible for the OTFS-ISAC system, while the NLOS channel cannot be perfectly sensed, resulting in a received signal that is a superposition of the known LOS and unknown NLOS signals. To demonstrate this phenomenon, the real channel fading $\bar h_{p}$ can be expressed as the sum of the true LOS channel $h_{p}$ and an estimated NLOS channel $\hat h_{p}$, given by:

\begin{align}
	\begin{split}
	    \bar h_{p}= h_{p}+ \hat h_{p},
	\end{split}
\end{align}
where $\hat h_p \sim \mathcal{C N}\left(0, e_{p}{|h_p|}^{2}\right)$ represents the NLOS channel, and \textcolor{black}{ $e_{p}$ denotes  the complexity of the environment obtained by the radar sensing}. A larger value of $e_{p}$ signifies the presence of more reflectors with higher reflection coefficients in the environment.

We introduce the notations $\bar R_{1}$ and $\bar R_{2}$ to represent the transmission achievable rates of users U1 and U2, respectively, when operating in an imperfect channel. The lower bound of $\bar R_{1}$ and $\bar R_{2}$ is established by considering the NLOS channel as interference. For user U1, the power of user U2, denoted by $E\left\{\omega_2\left(  {| h_{1}|}^{2}+    {| \hat h_{1}|}^{2}  \right)\right\}=\omega_{2}\left(\left|h_{1}\right|^{2}+e_{1}\left| h_{1}\right|^{2}\right) $, along with the NLOS component of user U1, denoted by  $E\left\{\omega_1 {| \hat h_{1}|}^{2} \right\}=\omega_1 e_{1} \left|h_{1}\right|^{2} $, are treated as noise. For user U2, the interference caused by the LOS component power of user U1 is removed through Successive Interference Cancellation (SIC), but the NLOS component power of user U1 still remains. Therefore, the NLOS component power of user U1, denoted by  $E\left\{\omega_1 {| \hat h_{2}|}^{2} \right\}=\omega_1 e_{2} \left|h_{2}\right|^{2} $ , and user U2, denoted by $E\left\{\omega_2 {| \hat h_{2}|}^{2} \right\}=\omega_2 e_{2} \left|h_{2}\right|^{2} $, are considered as noise for user U2. The lower bounds of the transmission achievable rates for U1 and U2 are expressed respectively as:
\begin{align}
	\begin{split}
	    \bar R_{1}^{\text{L}}=\log _{2}\left(1+\frac{\omega_{1}\left|h_{1}\right|^{2}}{\omega_{2}\left(\left|h_{1}\right|^{2}+e_{1}\left| h_{1}\right|^{2}\right)+\omega_1 e_{1}\left| h_{1}\right|^{2}+n_0}\right),
	\end{split}
\end{align}
and
\begin{align}
	\begin{split}
	    \bar R_{2}^{\text{L}}=\log _{2}\left(1+\frac{\omega_{2}\left|h_{2}\right|^{2}}{\omega_1 e_{2}\left| h_{2}\right|^{2}+\omega_2 e_{2}\left| h_{2}\right|^{2}+n_0}\right).
	\end{split}
\end{align}

Conversely, the upper bounds of $\bar R_{1}$ and $\bar R_{2}$ are obtained when the NLOS is leveraged for communication. The extra power to boost the rate is represented by $\omega_{1}e_{1}\left|h_{1}\right|^{2}$ and  $\omega_{2}e_{2}\left|h_{2}\right|^{2}$ for users U1 and U2, respectively. The upper bounds of the transmission achievable rates can be expressed respectively as:
\begin{align}
	\begin{split}
	    \bar R_{1}^{\text{U}}=\log _{2}\left(1+\frac{\omega_{1}\left(\left|h_{1}\right|^{2}+e_{1}\left| h_{1}\right|^{2}\right)}{\omega_{2}\left(\left|h_{1}\right|^{2}+e_{1}\left| h_{1}\right|^{2}\right)+n_0}\right), \label{R1Upper}
	\end{split}
\end{align}
and
\begin{align}
	\begin{split}
	    \bar R_{2}^{\text{U}}=\log _{2}\left(1+\frac{\omega_{2}\left(\left|h_{2}\right|^{2}+e_{2}\left| h_{2}\right|^{2}\right)}{\omega_1 e_{2}\left| h_{2}\right|^{2}+n_0}\right).
	\end{split} \label{R2Upper}
\end{align}

\subsubsection{MMF}
In the presence of an imperfect channel, the problem of optimizing the max-min fairness (MMF) becomes a constrained optimization problem, denoted by (P3), as:
\begin{align}
	\text{(P3)}:\mathop{\max\min}\limits_{\omega_1,\omega_2}\quad&\left\{\bar R_{1},\bar R_{2}\right\},\label{im_MMF} \\
	s.t.\quad&\omega_{1} + \omega_{2} \le P_{\text{t}}, \tag{\ref{im_MMF}{a}}
\end{align}
The objective function of (P3) is to maximize the minimum achievable rate, denoted by ${\bar R_{1}, \bar R_{2}}$. The constraint is that the sum of the power allocations for users U1 and U2 should not exceed the total transmit power $P_{\text{t}}$. When the lower bound performance of MMF is optimized, it is assumed that the channels for both users are highly correlated since U1 and U2 are in the same beam, i.e., $e_1 = e_2 = e$. In this case, the power allocation for user U2 can be expressed as $\omega_{2,\text{MMF}}^{{\text{L}},0}=\sqrt{P_{\text{t}}^{2}e^2+P_{\text{t}}^2e}-P_{\text{t}}e$ when the achievable rates for both users are equal, i.e., $\bar R_{1}^{\text{L}}=\bar R_{2}^{\text{L}}$. If $\omega_2 \ge \omega_{2,\text{MMF}}^{\text{L},0}$, the objective function of (P3) becomes ${\bar R_{1}^{\text{L}}}$, which increases as $\omega_2$ decreases. On the other hand, if $\omega_2\le\omega_{2,\text{MMF}}^{{\text{L}},0}$, the objective function of (P3) becomes ${\bar R_{2}^{\text{L}}}$, which increases as $\omega_2$ increases. Therefore, the optimal power allocation for users U1 and U2 in the lower bound of MMF is $\omega_{2,\text{MMF}}^{\text{L},*}=\omega_{2,\text{MMF}}^{\text{L},0}$ and $\omega_{1,\text{MMF}}^{\text{L},*}=P_{\text{t}}-\omega_{2,\text{MMF}}^{\text{L},0}$, respectively. The upper bound performance of the MMF optimization problem (P3) is investigated by assuming that the achievable rates for both users are equal, denoted by $\bar R_{1}^\text{U}=\bar R_{2}^\text{U}$. The optimal power allocation for user U2 in the upper bound of MMF is then obtained as $\omega_{2,\text{MMF}}^{\text{U},*}=\sqrt{1+e}-1$, and the optimal power allocation for user U1 is obtained as $\omega_{1,\text{MMF}}^{\text{U},*}=P_{\text{t}}-\omega_{2,\text{MMF}}^{\text{U},*}$, using a similar derivation as for the lower bound.
%%%%%%%%%%%%%%%%%
\subsubsection{Sum-Rate}

The SR optimization under imperfect channel can be formulated as:
\begin{align}
	(P4): \mathop {\max }\limits_{\omega_1,\omega_2} \quad 
	&\bar R_1+\bar R_2 ,\label{Y}\\
	\textrm{s.t} \quad 
	&\bar R_1\ge  R_{1,\text{min}},  \tag{\ref{Y}{a}}  \label{SR_C_a}\\
	&\bar R_2\ge  R_{2,\text{min}}, \tag{\ref{Y}{b}} \label{SR_C_b}\\
	&\omega_{1} + \omega_{2} \le P_{\text{t}}. \tag{\ref{Y}{c}}
\end{align}
The objective function of (P4) is to maximize the sum of achievable rates for users U1 and U2, denoted by $\bar R_1+\bar R_2$. The constraints of (P4) ensure that the achievable rates for both users are greater than or equal to a minimum rate requirement, denoted by $R_{1,\text{min}}$ and $R_{2,\text{min}}$, respectively. In addition, the total power allocated to users U1 and U2 should not exceed the total transmit power, denoted by $P_{\text{t}}$.

To investigate the lower bound performance of SR in (P4), we assume that the achievable rates for both users are equal to the lower bound of achievable rates, denoted by $\bar R_{1}^\text{L}$ and $\bar R_{2}^\text{L}$, respectively. The object function of (P4) for the lower bound can be expressed as ${f^{\text{L}}_{\text{SR}}}\left(\omega_2\right)=\bar R_1^{\text{L}}+\bar R_2^{\text{L}}$, where the power allocation for user U1 is  $\omega_1=P_{\text{t}}-\omega_2$. It is guaranteed that the corresponding derivative function ${f^{\text{L}}_{\text{SR}}}'\left(\omega_2\right)>0$ when $\omega_2\in [0,P_{\text{t}}]$. The optimal power allocation for user U2 in the lower bound of SR, denoted by $\omega_{2,\text{SR}}^{\text{L},*}$, can be obtained by finding the upper bound of $\omega_2$ that satisfies the constraints in \eqref{SR_C_a} and \eqref{SR_C_b}. The optimal power allocation for user U2 is then expressed as $\omega_{2,\text{SR}}^{\text{L},*}=\frac{P_{\text{t}}(1+e)+n_{0}/|h_{1}|^{2}}{2^{R_{1,\text{min}}}}-P_{\text{t}}e-n_{0}/|h_{1}|^2$, and the corresponding optimal power allocation for user U1 is $\omega_{1,\text{SR}}^{\text{L},*}=P_{\text{t}}-\omega_{2,\text{SR}}^{\text{L},*}$.

Then, in order to analyse the upper bound performance of the SR problem, denoted by (P4), we set $\bar R_{1}=\bar R_{1}^\text{U},~\bar R_{2}=\bar R_{2}^\text{U}$ in P(4). Using the power allocation $\omega_1=P_{\text{t}}-\omega_2$, the object function of (P4) is defined as $f_{\text{SR}}^{\text{U}}\left(\omega_2\right)=\bar R_1^{\text{U}}+\bar R_2^{\text{U}}$, which is further expressed as:

\begin{align}
	\begin{split}
f_{\text{SR}}^{\text{U}}\left(\omega_2\right)=&\log_{2}\left({\frac{P_{\text{t}}(1+e)+\frac{n_{0}}{|h_{1}|^2}}{P_{\text{t}}e+\omega_{2}+\frac{n_{0}}{|h_{1}|^2}}}\right)\\+&\log_{2}\left({\frac{P_{\text{t}}(e+2)-\omega_2+\frac{n_{0}}{|h_{2}|^2}}{P_{\text{t}}e+\frac{n_{0}}{|h_{2}|^2}}}\right)
	\end{split}
\end{align}
where the terms $\frac{n_{0}}{|h_{1}|^2}$ and $\frac{n_{0}}{|h_{2}|^2}$ can be ignored as they are very small compared to the others. Hence, the derivative function ${f_{\text{SR}}^{\text{U}}}^{'}\left(\omega_2\right)$ can be simply expressed:
\begin{align}
	\begin{split}
	  {f_{\text{SR}}^{\text{U}}}^{'}\left(\omega_2\right)=\frac{ {\omega_2}^{2}+2P_{\text{t}}e\omega_2-P_{\text{t}}^2}{\ln{2}\left(P_{\text{t}}e+\omega_2\right)\left(\omega_2P_{\text{t}}-\omega_2^2\right)}. 
	\end{split}\label{IM_Derivation}
\end{align}
% \begin{algorithm}[t]
% 	\renewcommand{\algorithmicrequire}{\textbf{Input:}} 
% 	\renewcommand{\algorithmicensure}{\textbf{Output:}}
% 	\caption{OTFS-ISAC Power allocation in the SR with NLOS}
% 	\footnotesize
% 	\begin{algorithmic}[1]
% 		\REQUIRE ~~\\
% 			Impact factor aoubt the NLOS $e$;\\
% 			Minimum data rate $R^{\rm{min}}_{1}$, $R^{\rm{min}}_{2}$;\\
% 			Total transmission power $P_{\rm{t}}$;\\
% 		\ENSURE ~~\\
% 		The power allocation  $\omega_1$ and $\omega_2$;\
% 		The sum rate of the system $\bar R_{1}^{U}+\bar R_{2}^{U}$;
% 		\STATE Calculate power allocation parameters Upper bound $\omega_2^{\text{ub}}$ and lower bound $\omega_2^{\text{lb}}$ through the limits;
% 		\IF{$\omega_2^{\text{lb}} \leq \omega_2^{\text{ub}}$} 
%             \STATE Return error;
%         \ELSE{}
%             \STATE Calculate optimal power allocation parameter $\omega_2^{SR_{0}^{u}}$ When the derivative of the objective function is 0;
%             \STATE Get the optimal power allocation parameter $\omega_2^{SR_{*}^{u}}$ by comparing the $\omega_2^{\text{lb}}$, $\omega_2^{\text{ub}}$ and $\omega_2^{SR_{0}^{u}}$;
%         \ENDIF 
% 		\STATE Calculate the Sum-Rate $R_{all}$;
% 	\end{algorithmic}
% \end{algorithm}
Observe from Eq.~\eqref{IM_Derivation},  the denominator of $ {f_{\text{SR}}^{\text{U}}}^{'}\left(\omega_2\right)$ is positive when $0\le \omega_2 \le P_{\text{t}}$. Furthermore, by setting the numerator to 0, the solution for the power allocation of user U2 can be obtained as $\omega_{2,\text{SR}}^{\text{U},0}=\sqrt{e^2P^2_{\text{t}}+P_{\text{t}}^2}-eP_{\text{t}}$. As a result, the function $f_2\left(\omega_2\right)$ decreases as $\omega_2$ increases  in the interval $\left(0,\omega_{2,\text{SR}}^{\text{U},0}\right)$ and increases as $\omega_2$ increases in the interval $\left(\omega_{2,\text{SR}}^{\text{U},0},P_{\text{t}}\right)$. Therefore, the optimal power allocation for user U2 in the upper bound of the SR problem is $\omega_{2,\text{SR}}^{\text{U},*}=\omega_{2,\text{SR}}^{\text{U},0}$ and the corresponding optimal power allocation for U1 is $\omega_{1,\text{SR}}^{\text{U},*}=P_{\text{t}}-\omega_{2,\text{SR}}^{\text{U},*}$. 
% For more clarity, the power allocation is shown in Algorithm 1.

% Obviously, the ${f}'\left(\omega_2\right)$ is positive under the premise of $h_1 \le h_2$. The upper and lower bounds of $\omega_2$ can be calculated through the limitation $\left(\ref{Y}{a}\right)$ and $\left(\ref{Y}{b}\right)$.
% \begin{align}
% 	\begin{split}
% 	  \le \omega_2 \le
% 	\end{split}
% \end{align}
% $\omega_2^u$ is the upper bound and the optimal solution.

\section{Numerical Results}
In this section, we provide the simulation results for our proposed NOMA-assisted OTFS-ISAC network with the aid of the proposed 3D motion prediction topology. Specifically, we evaluated the MMR and SR performance under perfect and imperfect channel conditions. The simulation parameters are summarized in table \ref{parametertable}.
\begin{table}[t]
	\centering
	\renewcommand{\arraystretch}{1.3} %设置行高
	\caption{Simulation Parameters}
	\label{table1} 
\begin{tabular}{c|c}
	\hline
	Parameter                & Value                 \\ \hline
	Carrier frequency (GHz) $f_c$  & 5                     \\ \hline
	OTFS frame size  $[M,N]$        & [1024,1024]           \\ \hline
	OTFS symbol duration (ms) $\Delta T$ & 4.4                   \\ \hline
	Transmit and receive gain (dB) $G_{\text{T}}$ and $G_{\text{R}}$  & 0                     \\ \hline
	UE speed (Kmph) $v$            & [30-60]               \\ \hline
    Guard frame size         & [30,60]               \\ \hline
    The distance of U1 and U2 (m) $[d_{1},d_{2}]$  & [7,15]             \\ \hline
    Channel estimation error $e$                 & [0-0.1]            \\ \hline
	%Channel        & Perfect \& Imperfect  \\ \hline
\end{tabular}\label{parametertable}
\end{table}

%1.Position

\begin{figure}[t]
		\centering
		\includegraphics[width=0.5\textwidth]{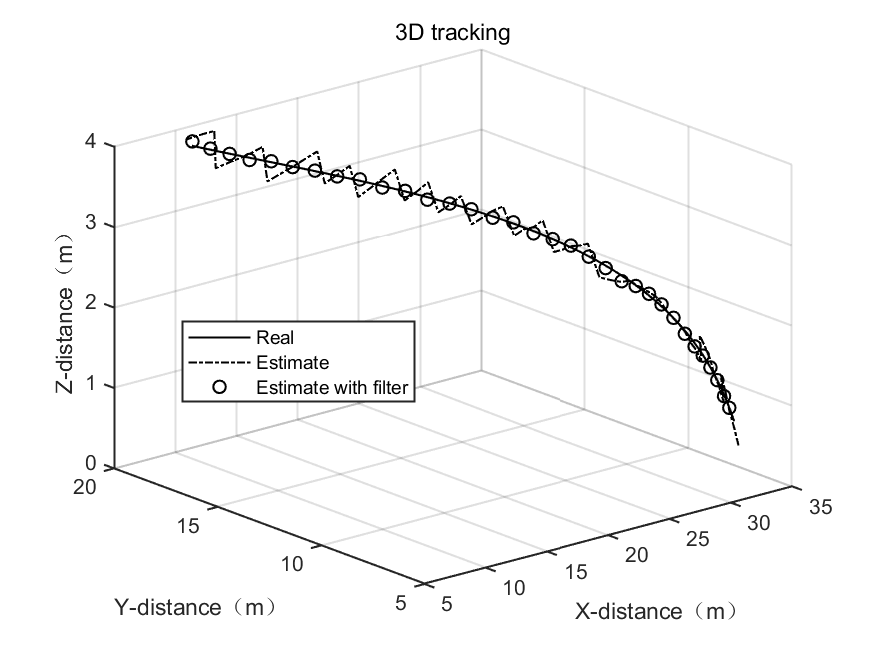}
		\caption{The 3D motion topology estimation.} \label{Speed}
\end{figure}
The performance of a 3D motion topological prediction system is demonstrated in Fig.~\ref{Speed}, where the system considers a user's movement along a curve with time-varying speed $v \in [9,13] ~\text{m/s}$. The solid line represents the user's actual movement, while the estimated position is illustrated by the dashed line. A low-pass filter with the method of moveing average  is employed to reduce the effect of the radar resolution-induced jitter on the user's continuous movement, thereby enhancing the accuracy of the position estimation. The proposed 3D motion topological approach successfully recovers the user's actual position, encompassing both azimuth and elevation information, with an estimation error of approximately 2\%, which fulfills the required accuracy level for user position tracking.

%2 MMF with noise 
\begin{figure}[t]
		\centering
		\includegraphics[width=0.5\textwidth]{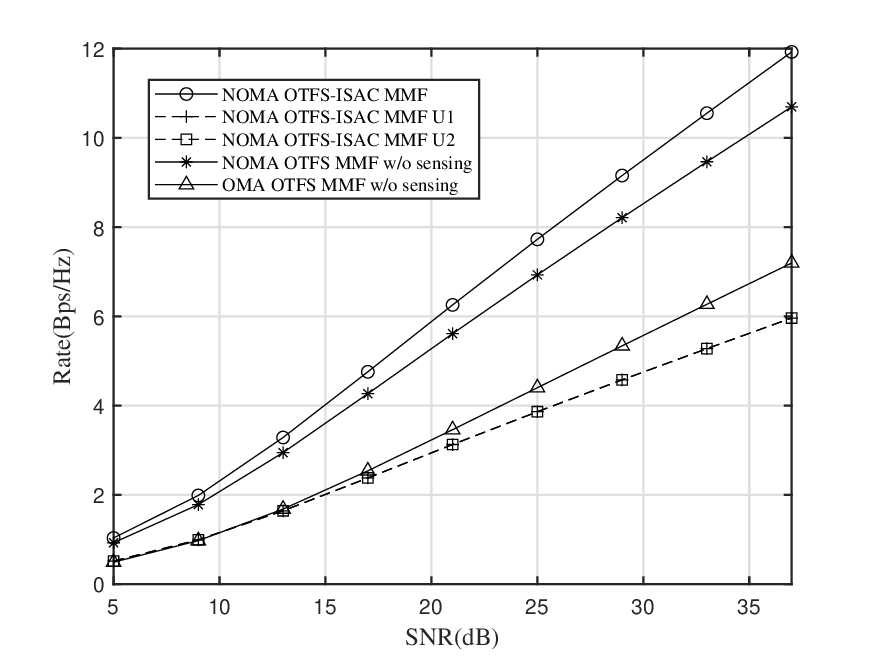}
		\caption{The comparison between different systems in the MMF problem over the perfect channel.} \label{Kvalue}\label{MMF_pic}
\end{figure}
Fig.~\ref{MMF_pic} depicts the achievable rate performance of MMF, assuming perfect channel conditions. The performance of three transmission protocols, namely NOMA-assisted OTFS-ISAC, NOMA-assisted OTFS without sensing, and OMA-assisted OTFS without sensing, are compared under varying values of SNR. Our proposed system outperforms the other systems, as evidenced by its highest achievable rate. \textcolor{black}{The NOMA-assisted version, which enables the spectrum to be shared among different users, yields higher spectral efficiency. The sensing can reduce the pilot overhead, which result in more information can be transmitted in the DD-domain. The objective function of (P1) ensures fairness between U1 and U2, resulting in both users having a rate that is half of the overall rate under different SNR values.}

%3 SR with noise 
\begin{figure}[t]
		\centering
		\includegraphics[width=0.5\textwidth]{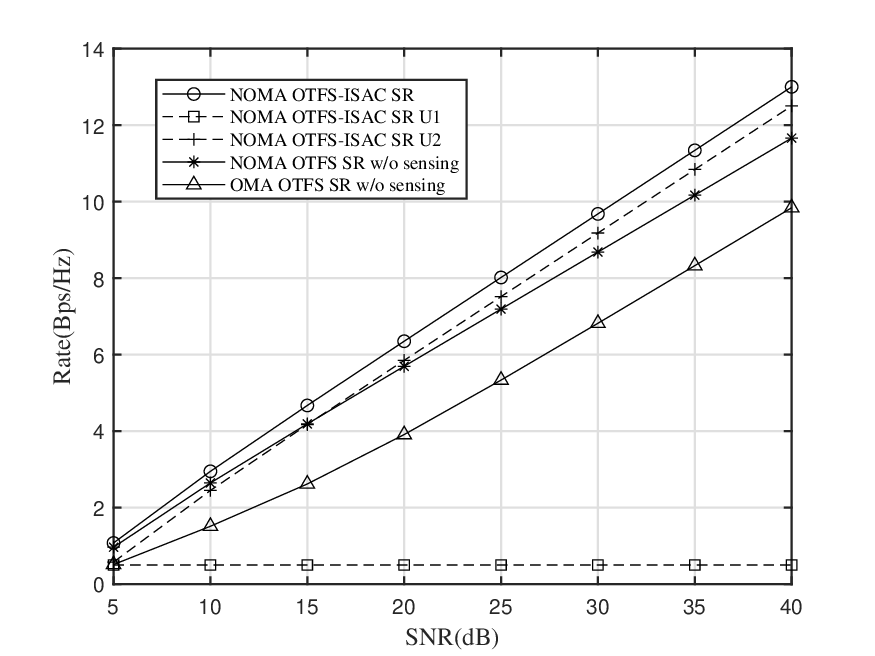}
		\caption{The comparison between different systems in the  SR problem  over the perfect channel.} \label{SR_pic}
\end{figure}
Fig.~\ref{SR_pic} presents the performance of the SR problem in the perfect channel scenario. The proposed NOMA-assisted OTFS-ISAC system, leveraging the benefits of both NOMA and sensing, achieves the highest rate compared to other techniques, consistent with the conclusion of the MMF problem, as depicted in Fig.\ref{MMF_pic}. However, \textcolor{black}{the MMF problem fairly satisfies  information transmission for multiple users, the SR problem focus  demonstrating  the overall performance of the ISAC system, whihc aims to maximize the sum rate.} Specifically, the system prioritizes increasing the rate of user U2 with the superior channel, while satisfying the minimum rate requirement of user U1 (0.5Bps/Hz). The data rate of user U2 increases with the SNR, surpassing that of user U1.

%4. MMF with the imperfect NLOS
\begin{figure}[t]
		\centering
		\includegraphics[width=0.5\textwidth]{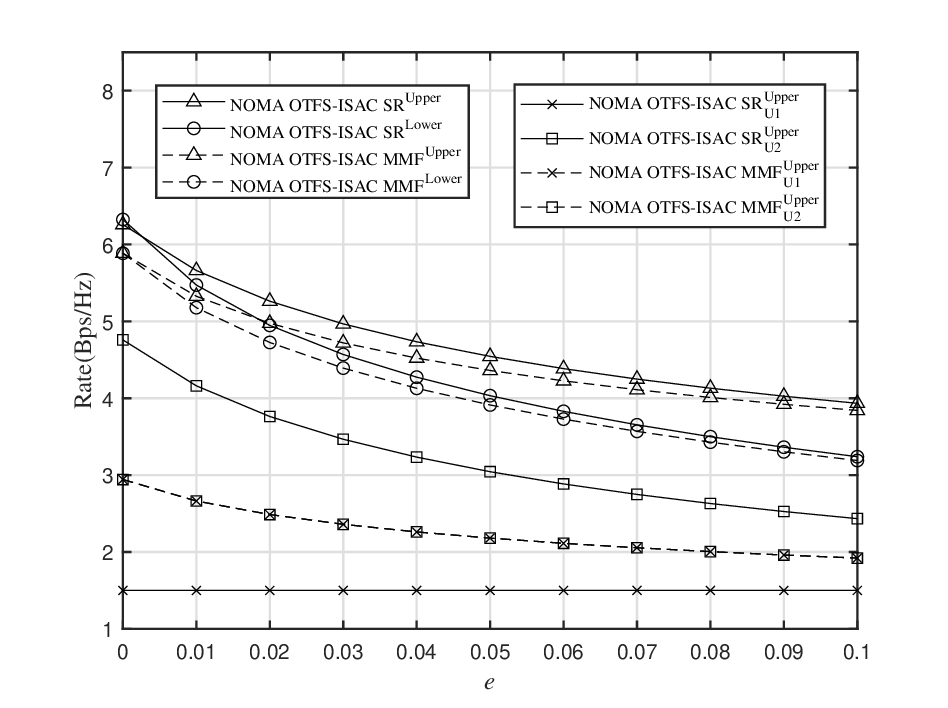}		\caption{The upper bound and lower bound performance of MMF and SR over the imperfect channel.} \label{SR_MMF_im}
\end{figure}
To demonstrate the impact of imperfect channel conditions on the system, we depict the extremities—both upper and lower—of MMF and SR against channel estimation inaccuracies, denoted as $e\in [0,0.1]$, in Fig.~\ref{SR_MMF_im}, which is consistent with the range of parameter assumptions for the Rice channel.  The upper boundary is derived by interpreting the NLOS power as a distinct gain, whereas the lower demarcation perceives it as interference. Notably, even when the NLOS power is viewed as an isolated gain for the upper threshold, it concurrently introduces interference for the alternative user within the system. This intrinsic relationship is described by the equations Eq.~\eqref{R1Upper} and Eq.~\eqref{R2Upper}. As $e$ increases from $0$ to $0.1$,  the MMF and SR rates manifest a pronounced deterioration. 
 A diminutive $e$ corresponds to closely spaced upper and lower thresholds for both SR and MMF rates. In scenarios devoid of NLOS (where $e=0$), these thresholds converge. The ascent of $e$ instigates a more pronounced descent in the lower threshold relative to its upper counterpart.Regarding the SR upper boundary, user U1 consistently registers a rate of 1.5~Bps/Hz, sustaining the baseline rate threshold with growing $e$. Conversely, the rate for user U2 exhibits a decrement with the escalation in $e$. Within the MMF upper bound, the rates of users U1 and U2 are equal to ensure fairness. These observations validate the precision of our antecedent NOMA-integrated OTFS-ISAC power distribution approach for both SR and MMF, particularly when accommodating imprecise channel conditions.
\begin{figure}[t]
		\centering
		\includegraphics[width=0.5\textwidth]{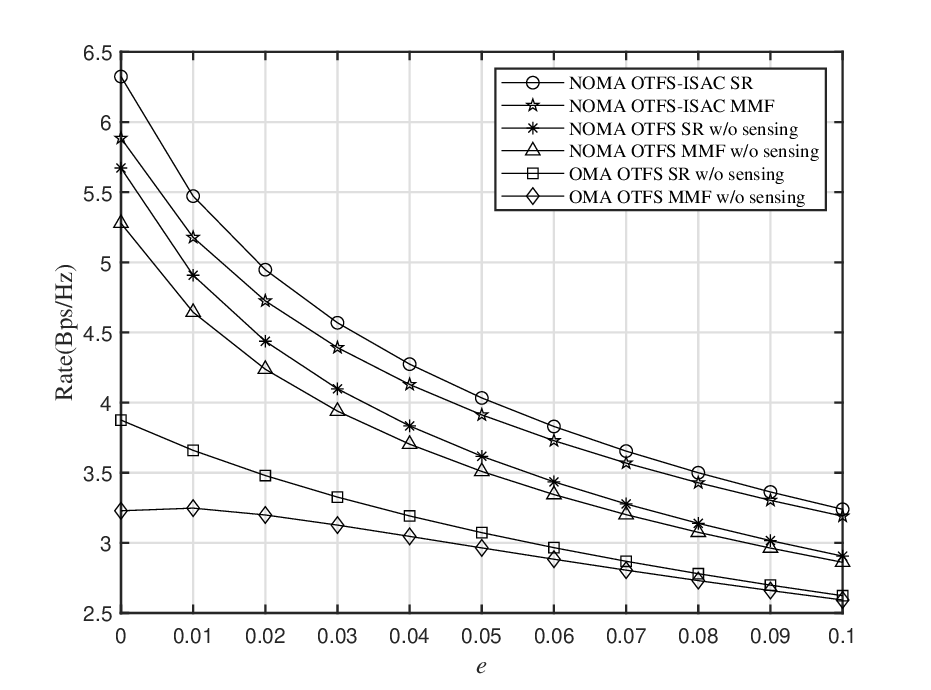}
		\caption{The comparison between different system  in MMF and SR problem over the imperfect channel.} \label{MMF_SR_im}
\end{figure}

Fig.\ref{MMF_SR_im} presents the evaluation of the proposed system's superiority over \textcolor{black}{other counterparts without sensing} under imperfect channel estimation. %The NOMA-assisted OTFS without sensing and the OMA version are considered as the benchmark, similar to the perfect channel scenario. 
The results indicate that the NOMA-assisted OTFS-ISAC system outperforms the benchmark by leveraging the benefits of NOMA and sensing, as discussed in Fig.\ref{MMF_pic}. To ensure fairness in the MMF problem, more power is allocated to U1, despite having a worse channel. It is observed that the system's rate considering the SR is higher than that considering the MMF, as $e$ increases from 0 to 0.1.

% \begin{figure}[t]
% 		\centering
% 		\includegraphics[width=80mm]{figure/SNR_K.eps}
% 		\caption{The comparison between MMF and SR with different $e$ and SNR.} \label{SNR_K}
% \end{figure}
% In Fig.~\ref{SNR_K} We evaluate the impact caused by the $e$ and SNR for the MMF and SR problem. The performance of the system with the smaller value $e=0.05$ is better than that with $e=0.1$, because the $e=0.05$ implies smaller NLOS power interference. Due to the impairment of the NLOS, when the SNR is higher than 30 dB, its impact on the system rate is in a low level. 

\begin{figure}[t]
		\centering
		\includegraphics[width=0.52\textwidth]{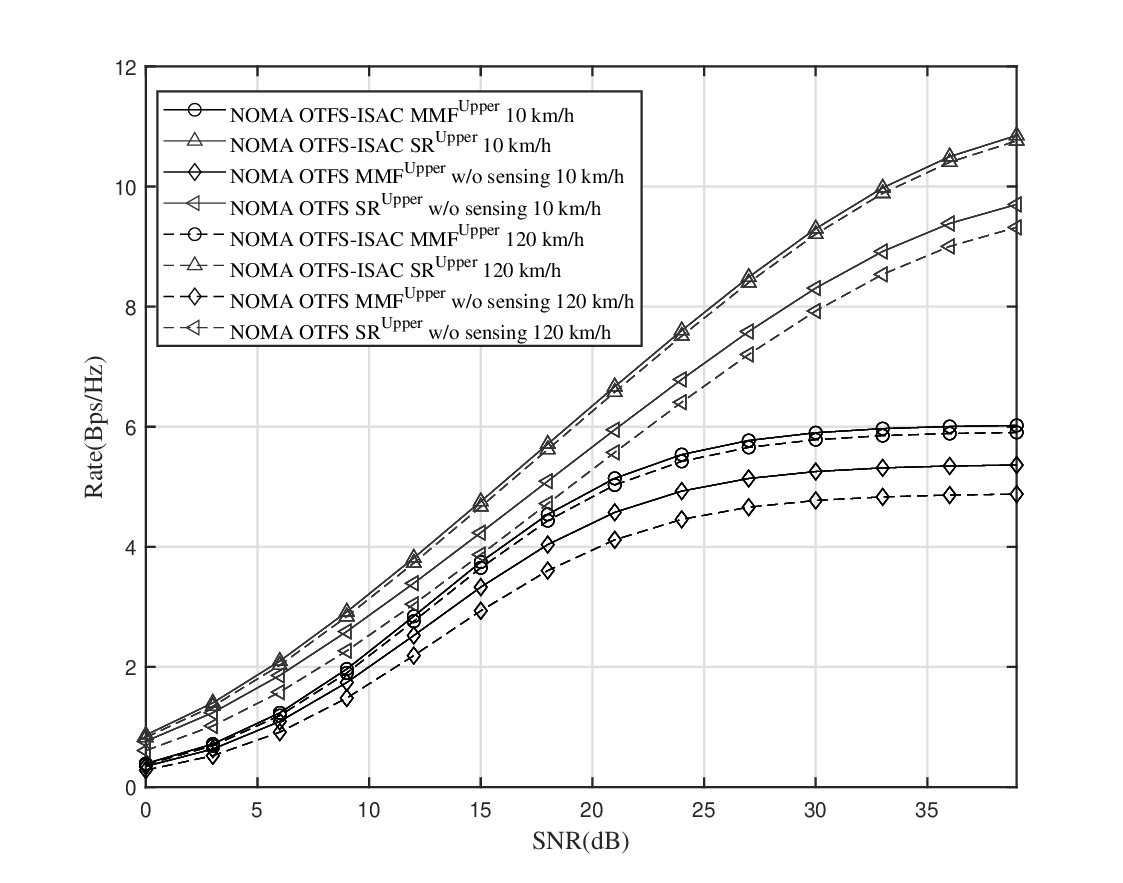}
		\caption{The comparison between MMF and SR  under different speed with $e=0.02$.}\label{Speed_d}
\end{figure}

Finally, the impact of speed on the system illustrated in Fig.~\ref{Speed_d} is investigated. Observations reveal that system performance decreases as the speed of the system increases. This decrease in performance is attributed to the widening of the position gap between the actual value and estimation of the system without sensing due to the higher speed. Moreover, the performance of the NOMA-OTFS system without sensing experiences a higher degradation. However, the incorporation of real-time motion prediction in the NOMA-assisted OTFS-ISAC results in a smaller degradation in performance. Additionally, the performance degradation in the presence of NLOS is less significant when the SNR exceeds 30 dB.

\section{Conclusion}
In this paper, we proposed a novel NOMA-assisted OTFS-ISAC network, where a UAV serves as an air base station to support multiple users. The system employs the OTFS waveform to extract the user's position and velocity information from the echo signals during communication. A three-dimensional motion model is proposed to retrieve the distance, velocity, and angle information of   users from the echo signals. The impact of the NLOS channel on the robust power allocation is evaluated for two NOMA classic problems: maximum SR and MMF. The proposed NOMA-assisted OTFS-ISAC system is demonstrated to achieve superior achievable data performance over the benchmark systems in terms of SR and MMF under both perfect and imperfect channel assumptions. 
	
	\bibliography{refer} 
\end{document}